\documentclass{moriond}

\def\Journal#1#2#3#4{{#1} {\bf #2}, #3 (#4)}

\def\PRL{\em Phys. Rev. Lett.}
\def\PRD{{\em Phys. Rev.} D}

\def\be{\begin{equation}}
\def\ee{\end{equation}}
\def\bea{\begin{eqnarray}}
\def\eea{\end{eqnarray}}

\newcommand{\Photo}{}

\begin{document}

\begin{flushright}
 \hspace{3cm} MS-TP-18-14
\end{flushright}

\title{Constraining PDFs from neutral current Drell-Yan measurements and \\
effects of resummation in slepton pair production}

\author{Juri Fiaschi}
\address{Institut f\"ur Theoretische Physik, Westf\"alische Wilhelms-Universit\"at M\"unster, Wilhelm-Klemm-Stra\ss{}e 9, D-48149 M\"unster, Germany}

\author{Elena Accomando}
\address{School of Physics \& Astronomy, University of Southampton, Highfield, Southampton SO17 1BJ, UK}

\author{Francesco Hautmann}
\address{Theoretical Physics Department, University of Oxford, Oxford OX1 3NP, UK}

\author{Michael Klasen}
\address{Institut f\"ur Theoretische Physik, Westf\"alische Wilhelms-Universit\"at M\"unster, Wilhelm-Klemm-Stra\ss{}e 9, D-48149 M\"unster, Germany}

\author{Stefano Moretti}
\address{School of Physics \& Astronomy, University of Southampton, Highfield, Southampton SO17 1BJ, UK}

\maketitle\abstracts{
The high statistics that will be collected during the LHC Run-II (and beyond) open the path to precision measurements at the TeV scale, where the PDFs will play a crucial role in BSM searches.
In the di-lepton final state accurate measurements of the Forward-Backward Asymmetry (AFB) will be available, particularly in the invariant mass region around the Z peak.
We show that this observable displays a statistical error which is competitive with that assigned to the existing PDF sets and which will rapidly become smaller than the latter as the integrated luminosity grows,
thereby offering a means of constraining the (anti)quark PDFs over a sizeable $(x,Q^2)$ range.
In the context of SUSY searches we consider the effects of employing threshold-improved PDFs in a consistent calculation at NLO+NLL of slepton pair production cross sections.
The calculations featuring a consistent resummation procedure both at PDF and partonic matrix element level are accompanied by PDF and scale uncertainties, and they provide a reliable and updated theoretical estimation for
experimental data analyses at the LHC Run-II.
}

\section{Introduction}

The LHC programme has recently entered in the Run-II stage, featuring an upgraded c.o.m. energy of 13 TeV and aiming to achieve a high integrated luminosity in the next years
and furthermore during the following High Luminosity (HL) stage.
In order to keep up with the increasing statistical precision of experimental measurements,
an impressive effort has been made on the theoretical side to provide higher order calculations, often including also the resummation of large logarithmic contributions that appear in the perturbative expansion,
such that in many cases the remaining uncertainty is dominated by the determination of the Parton Distribution Functions (PDFs).
In this context, a precise determination of the PDFs will be a crucial point for LHC physics, as well as their consistent employment in the theoretical calculations of cross sections
at next-to-leading order (NLO) and next-to-leading logarithmic (NLL) accuracy.

These two main points are considered in this work.
In Sect.~\ref{sec:AFB}, we propose the inclusion of the di-lepton final state Forward-Backward Asymmetry (AFB) pertaining to the Neutral Current (NC) Drell-Yan (DY) production channel, in the fit of the PDFs.
In Sect.~\ref{sec:resummation} we consider a SUSY BSM scenario and we study the effect of adopting threshold-resummed improved PDFs in a consistent calculation of the slepton pair production cross section at NLO+NLL.

\section{The Forward-Backward Asymmetry in the fit of PDFs}
\label{sec:AFB}

\begin{figure}[h]
\begin{center}
\includegraphics[width=.30\textwidth]{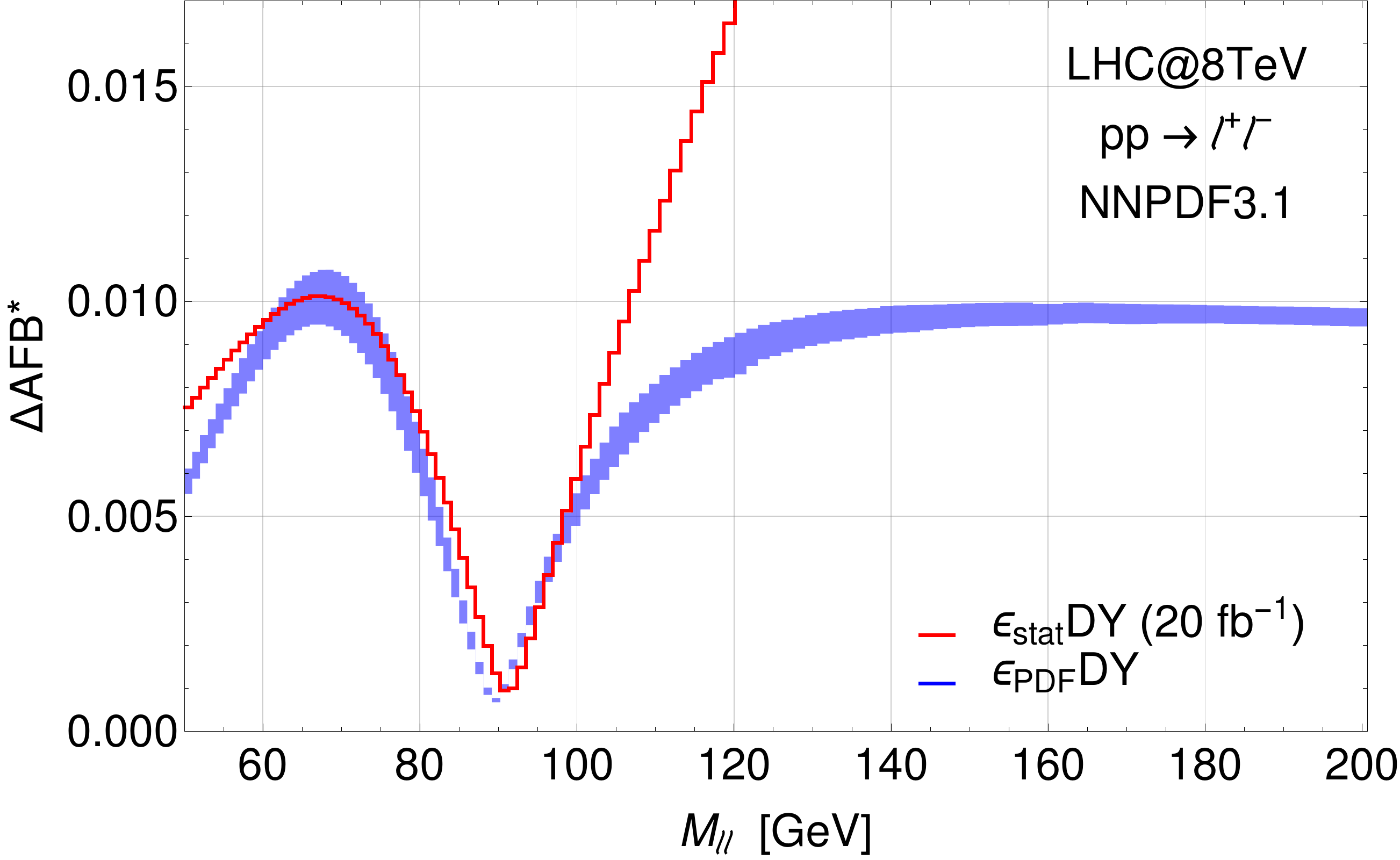}
\includegraphics[width=.30\textwidth]{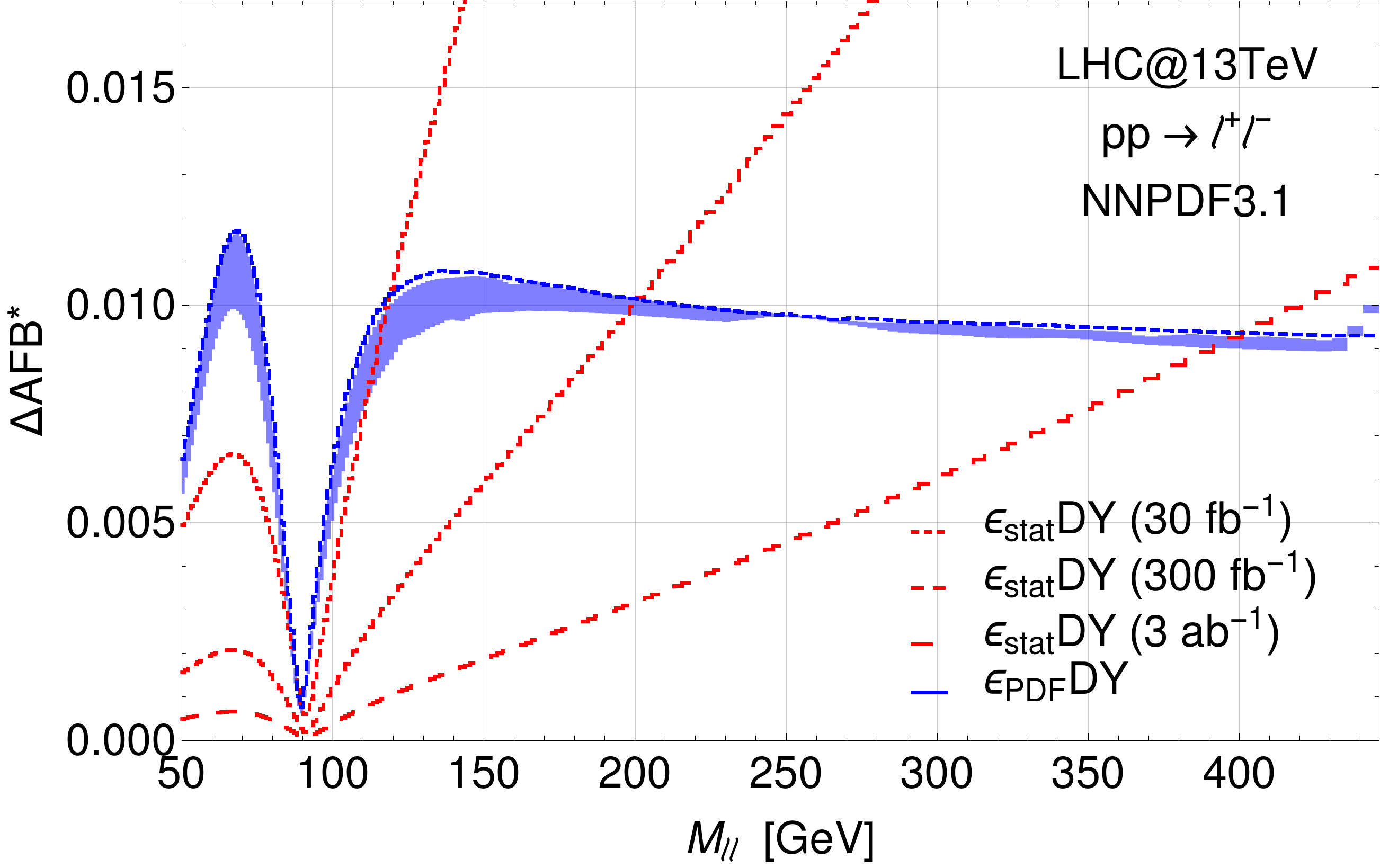}
\caption{Statistical and PDF uncertainties on the reconstructed $A_{\rm FB}^*$ distribution for the LHC with c.o.m. energy of 8 TeV (left) and 13 TeV (right).
The blue band refers to the PDF uncertainty evaluated varying the factorisation scale in the interval $0.5 M_{\ell\ell} < Q  < 2 M_{\ell\ell}$, while the dashed blue line (right plot only) represent the choice $Q  = p_T$.
The statistical error is obtained for different integrated luminosities as specified in the legend.
}
\label{fig:AFB_8_13_TeV}
\end{center}
\end{figure}

In this section we consider the possibility of including the Forward-Backward Asymmetry (AFB) observable in future fits of the PDFs~\cite{Accomando:2017scx}.
In Fig.~\ref{fig:AFB_8_13_TeV} we compare the statistical and PDF errors on the AFB and we recognise that when the former is smaller than the latter,
a precise experimental measurement of the observable can improve the fit of the PDFs, thus reducing its PDF uncertainty.
In the plot on the left are shown the two sources of uncertainties in the Run-I setup.
As visible the statistical indetermination is always of the same order or larger than the PDF error,
thus no improvement in the PDF fit is to be expected due to the inclusion of this data.
In the plot on the right the same exercise is repeated for the Run-II c.o.m. energy and for various stages of achieved integrated luminosity.
In this scenario it is possible to define an invariant mass interval where the statistical precision overcomes the PDF error.

Another interesting feature of the AFB resides in its sensitivity on the partonic content of the proton which is parametrised differently in each PDF set.
The AFB indeed depends on the relative size of the $u$ and $d$ quarks contribution to the DY process.
The differences in the parametrisation of the quarks PDFs between the various sets is augmented in the high-$x$ region.
Imposing a rapidity cut on the sample we select the di-lepton events arising with a large boost, thus originating from the interaction between a valence quark with a large-$x$ and a sea anti-quarks with small-$x$.

\begin{figure}[h]
\begin{center}
\includegraphics[width=.30\textwidth]{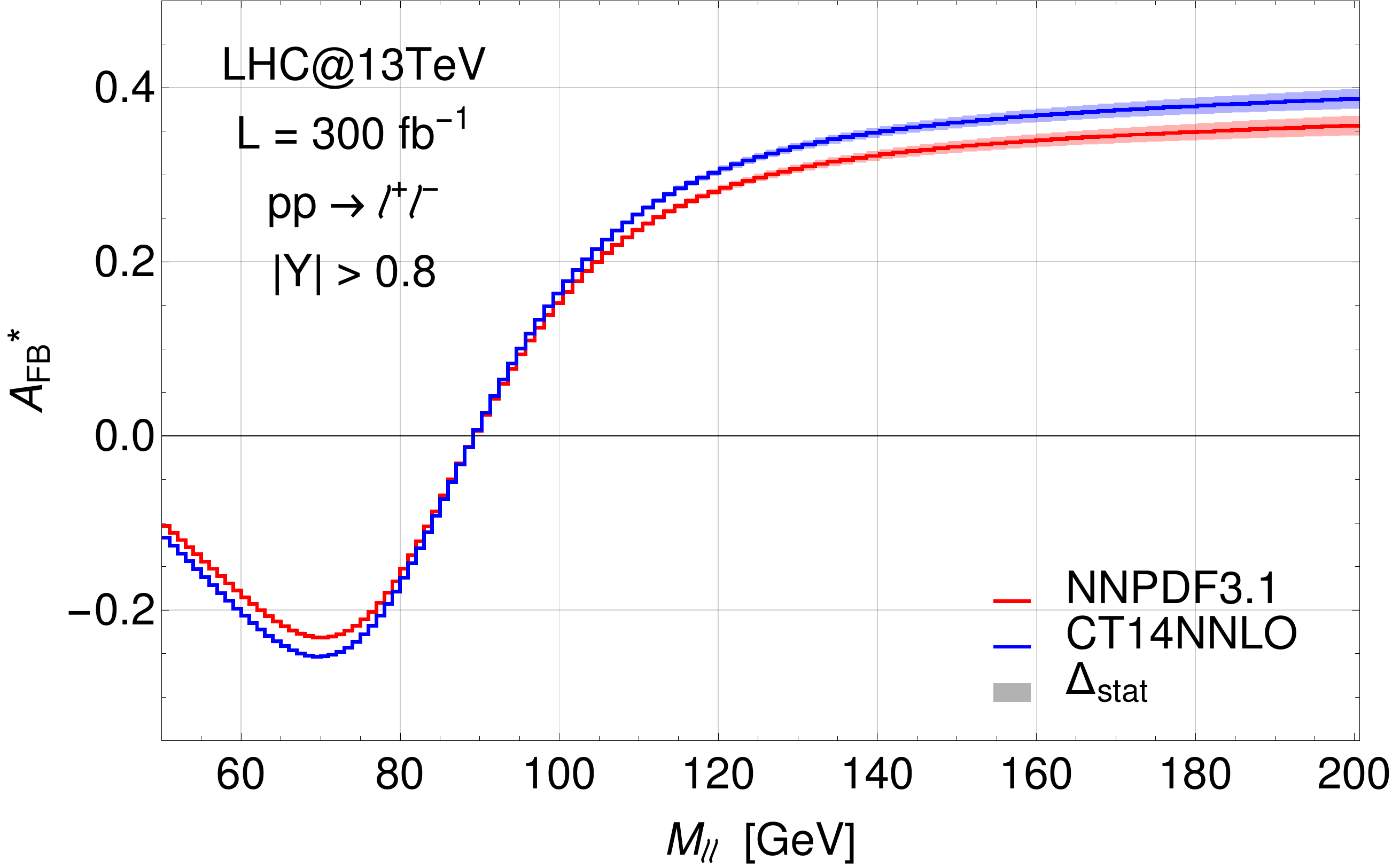}
\includegraphics[width=.30\textwidth]{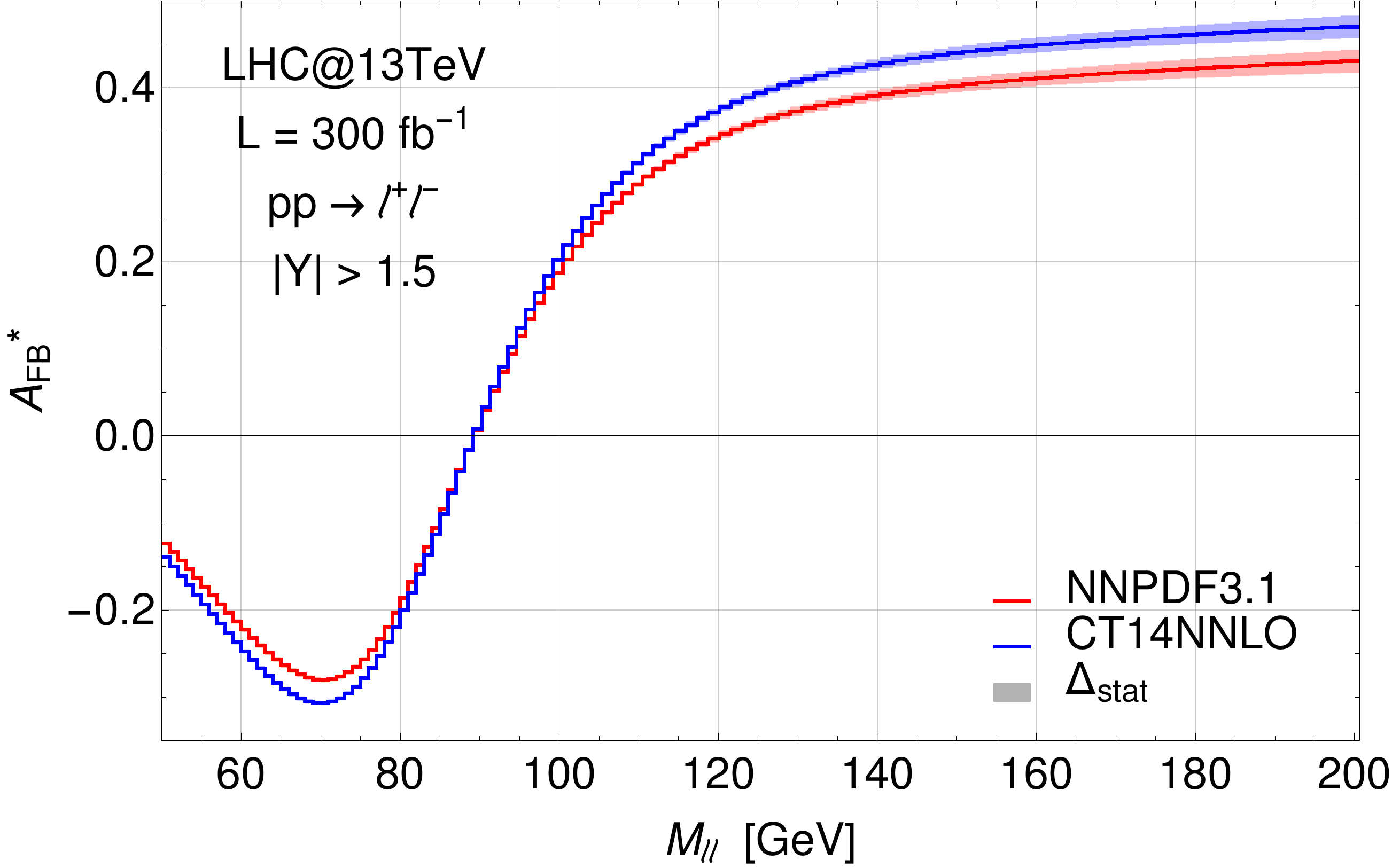}
\caption{$A_{\rm FB}^*$ distribution in the invariant mass region around the $Z$ peak at the LHC with c.o.m. energy of 13 TeV computed with the CT14NNLO and the NNPDF3.1 PDF sets.
The error band represent the statistical uncertainty computed for an integrated luminosity $L = 300$ fb$^{-1}$.
A rapidity cut of $|Y_{\ell\ell}| > 0.8$ (left) and $|Y_{\ell\ell}| > 1.5$ (right) is imposed on the di-lepton system.
}
\label{fig:AFB_Y_cut}
\end{center}
\end{figure}

In Fig.~\ref{fig:AFB_Y_cut} we are showing the predictions for the AFB obtained with the NNPDF3.1~\cite{Ball:2017nwa} and the CT14NNLO~\cite{Dulat:2015mca} PDF sets.
From left to right, we have applied a rapidity cut $|Y| > 0.8$ and $|Y| > 1.5$ on the observable, while the statistical uncertainty bands have been computed for an integrated luminosity $L = 300$ fb$^{-1}$.
As visible the separation between the predictions of the two PDF sets grows with the rapidity cut, thus an experimental measurement of the AFB in these conditions can be used to resolve the correct parametrisation
of the quark (anti-quarks) PDFs in the high-$x$ (low-$x$) region and to extract information on the $u$ and $d$ quarks content in the proton.

\section{Threshold-resummed improved PDFs in slepton pair production cross sections}
\label{sec:resummation}

Searches for BSM physics will reach considerably higher sensitivity as the LHC machine will operate at higher c.o.m. energy and luminosity.
Impressive theoretical efforts have been spent to keep up with the increasing precision of experimental measurements, with the aim of providing reliable predictions for the relevant processes in the experimental analysis.
This holds also in the context of SUSY searches since nowadays the cross sections for many processes have been calculated at NLO and beyond~\cite{Beenakker:1999xh}.
Also special resummation techniques have been established in order to take into account the contribution of logarithmic terms that appear in the perturbative expansions of the cross sections to all orders.
In this work we make use the public code RESUMMINO~\cite{Fuks:2013vua}, which has been developed specifically for the calculation of resummed cross sections at NLL precision for several SUSY processes,
with the purpose of updating the theoretical predictions for slepton pair production cross sections with NLO+NLL accuracy for the c.o.m. energy of the LHC Run-II~\cite{Fiaschi:2018xdm}.

A consistent calculation of the cross sections requires that the order in the perturbation expansion of the partonic matrix element of the hard process matches the one of the fit of the PDFs which are employed in the computation.
For this reason we will adopt the threshold-resummation improved PDFs provided by the NNPDF collaboration~\cite{Bonvini:2015ira}.
This particular set includes grids that have been obtained using matrix elements calculated at NLO (NNPDF30\_nlo\_disdytop) and NLO+NLL (NNPDF30\_nll\_disdytop) in the fit of a reduced experimental data set, including only 
Deep Inelastic Scattering (DIS), Drell-Yan and top pair production data.
Consequently to this reduction these PDF sets are generally affected by a larger error with respect to the case of globally fitted PDFs.

We present our results in the form of a $K$-factor, following the prescription of Ref.~\cite{Beenakker:2015rna}, which is defined as:
\begin{equation}
 K =
 \frac{\sigma({\rm NLO+NLL})_{\rm NLO~global}}{\sigma({\rm NLO})_{\rm NLO~global}}
 \cdot
 \frac{\sigma({\rm NLO+NLL})_{\rm NLO+NLL~reduced}}{\sigma({\rm NLO+NLL})_{\rm NLO~reduced}},
 \label{eq:K_factor}
\end{equation}
The purpose of this choice is two-fold.
On one hand this definition allows to obtain (approximate) central total NLO+NLL cross sections with NLO+NLL PDFs via
\begin{equation}
 \sigma({\rm NLO+NLL})_{\rm NLL+NLO~global} = K \cdot \sigma({\rm NLO})_{\rm NLO~global}.
 \label{eq:K_application}
\end{equation}
On the other hand we can rescale the (smaller) PDF error obtained from the global PDF set directly on the $K$-factor
such that it will be straightforward to estimate the uncertainty on the consistent result at NLO+NLL.
However in order to keep an important benefit of the resummation we will transfer on the $K$-factor the relative size of the scale uncertainty calculated on the NLO+NLL result.
We will also sum in quadrature the two (independent) sources of uncertainty in order to obtain an overall theoretical error band.

\begin{figure}[h]
\begin{center}
\includegraphics[width=.30\textwidth]{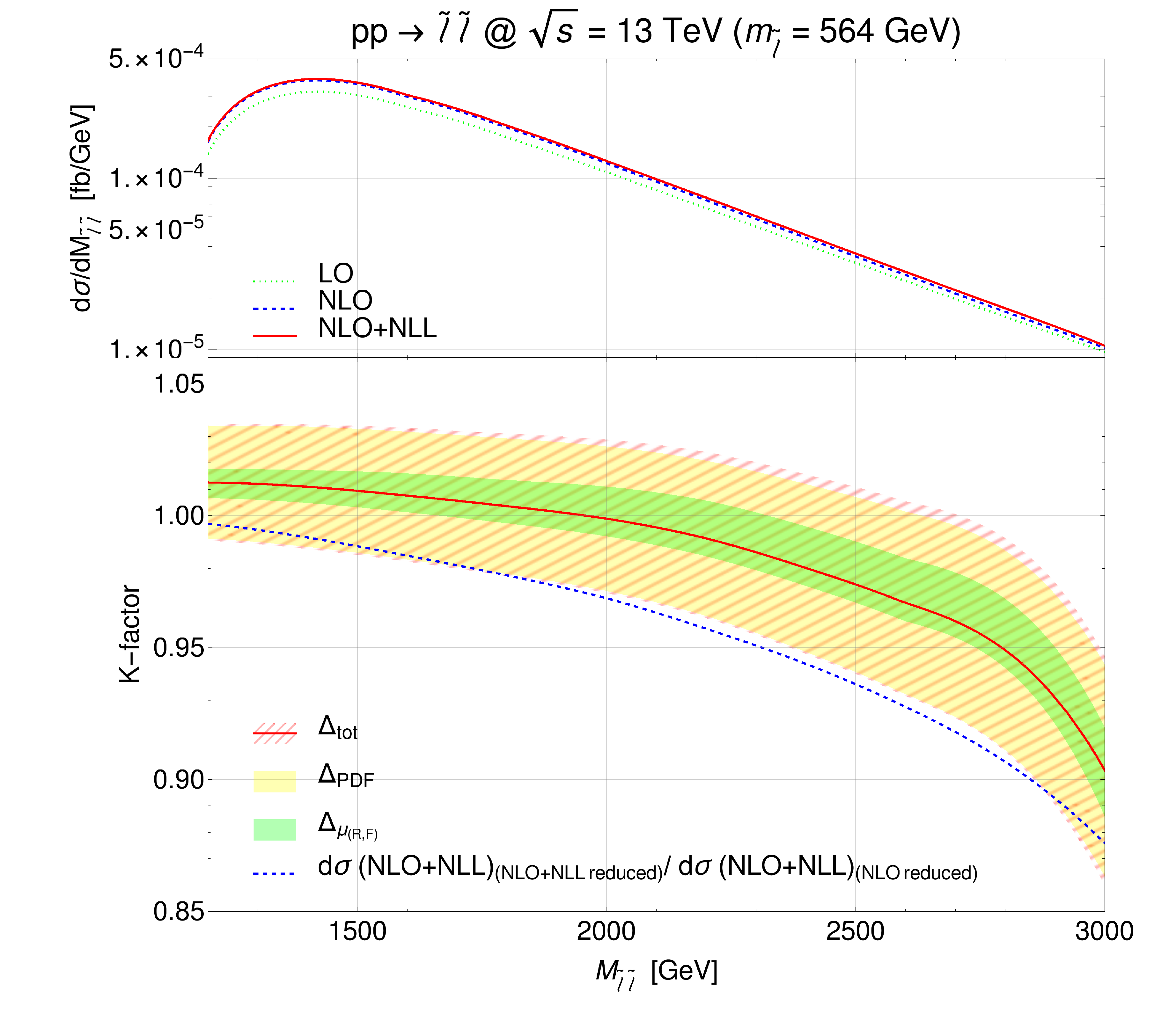}
\includegraphics[width=.30\textwidth]{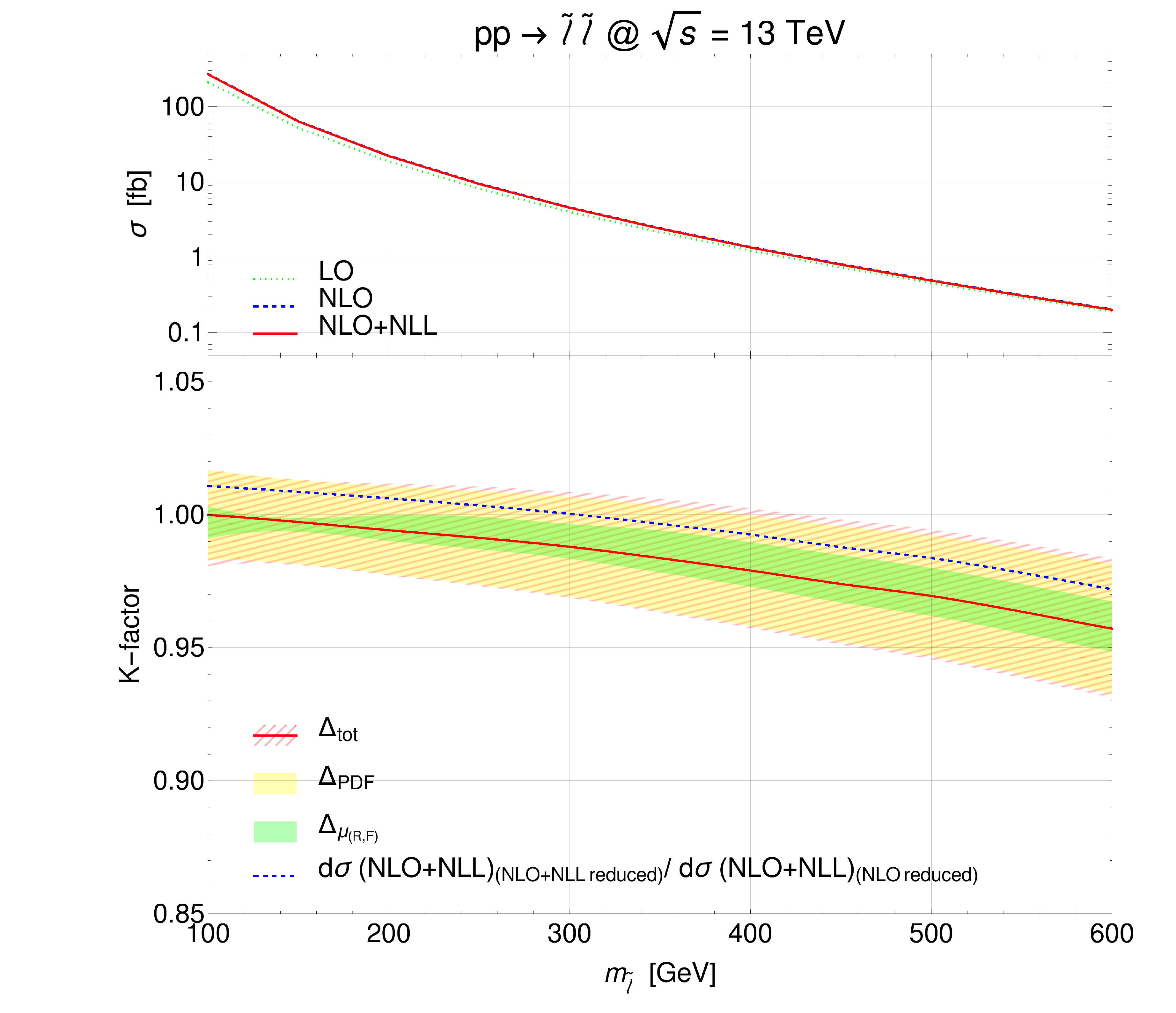}
\caption{
Invariant-mass (left) and integrated cross section (right) distributions with their $K$-factors (lower panels) according to Eq.(\ref{eq:K_factor})
using the full expression (full red) and only its second, PDF-dependent part (dashed blue line)
for the pair production of left-handed selectrons/smuons at the LHC with $\sqrt{s} = 13$ TeV.
}
\label{fig:first_generation}
\end{center}
\end{figure}

In Fig.~\ref{fig:first_generation} we show the results for the differential (left) and integrated (right) cross section of first and second generation slepton pair production.
The differential cross section has been evaluated for one choice of the SUSY parameters which predicts a slepton mass of 564 GeV,
while the total cross section has been obtained for a range of slepton masses within the sensitivity of the LHC by the end of the Run-II~\cite{CMS-PAS-SUS-17-009}.
In the lower plots we show the $K$-factor as defined in Eq.~\ref{eq:K_factor} (red curve) and the ratio of the central values obtained with the reduced PDF sets fitted at NLO and NLO+NLL (blue dashed line).
The latter highlights the effect of the resummation in the sole PDF fit.
Comparing these two curves we notice a partial compensation between the effect of the resummation within the PDFs and the partonic matrix element.

\begin{figure}[h]
\begin{center}
\includegraphics[width=.30\textwidth]{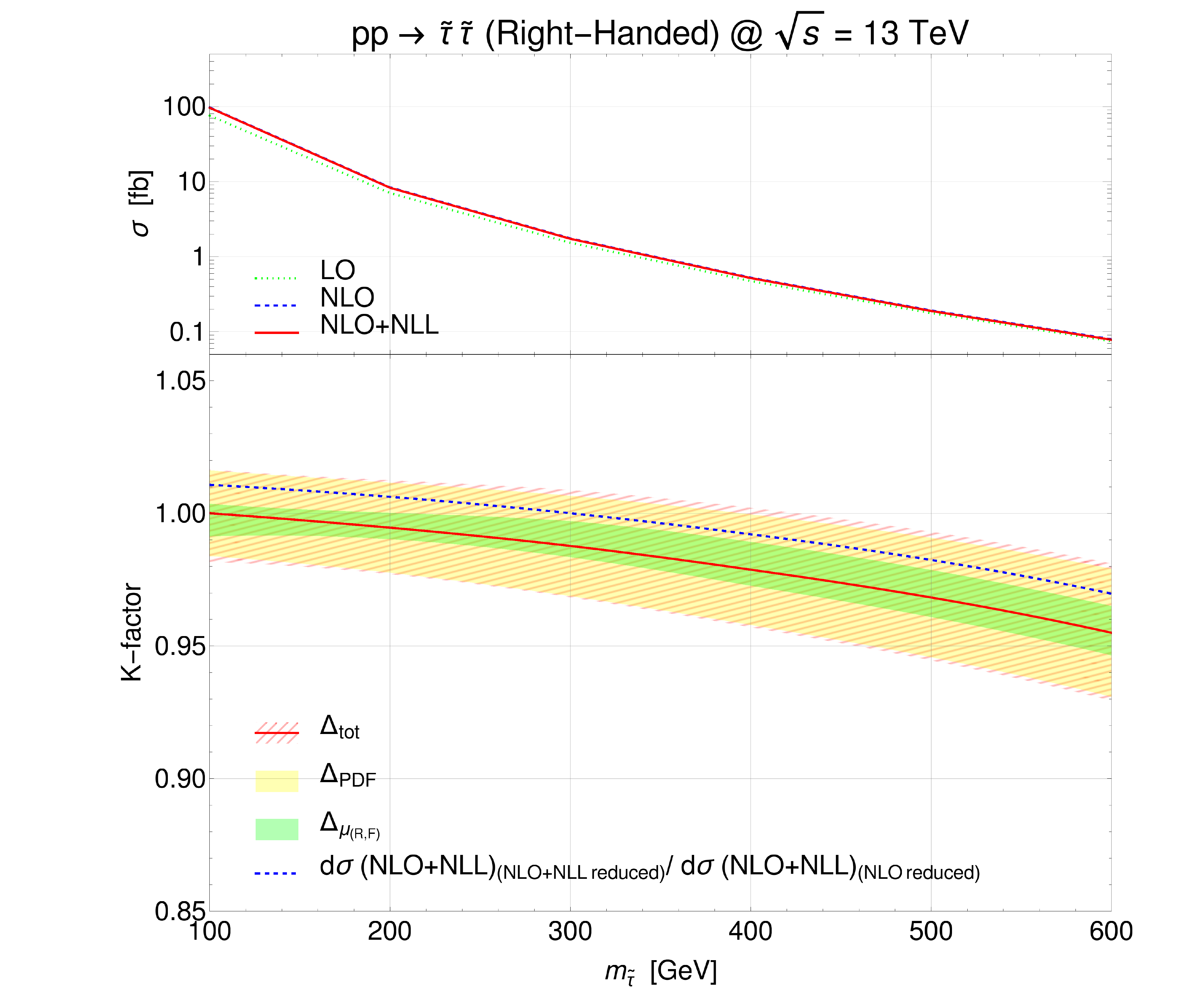}
\includegraphics[width=.30\textwidth]{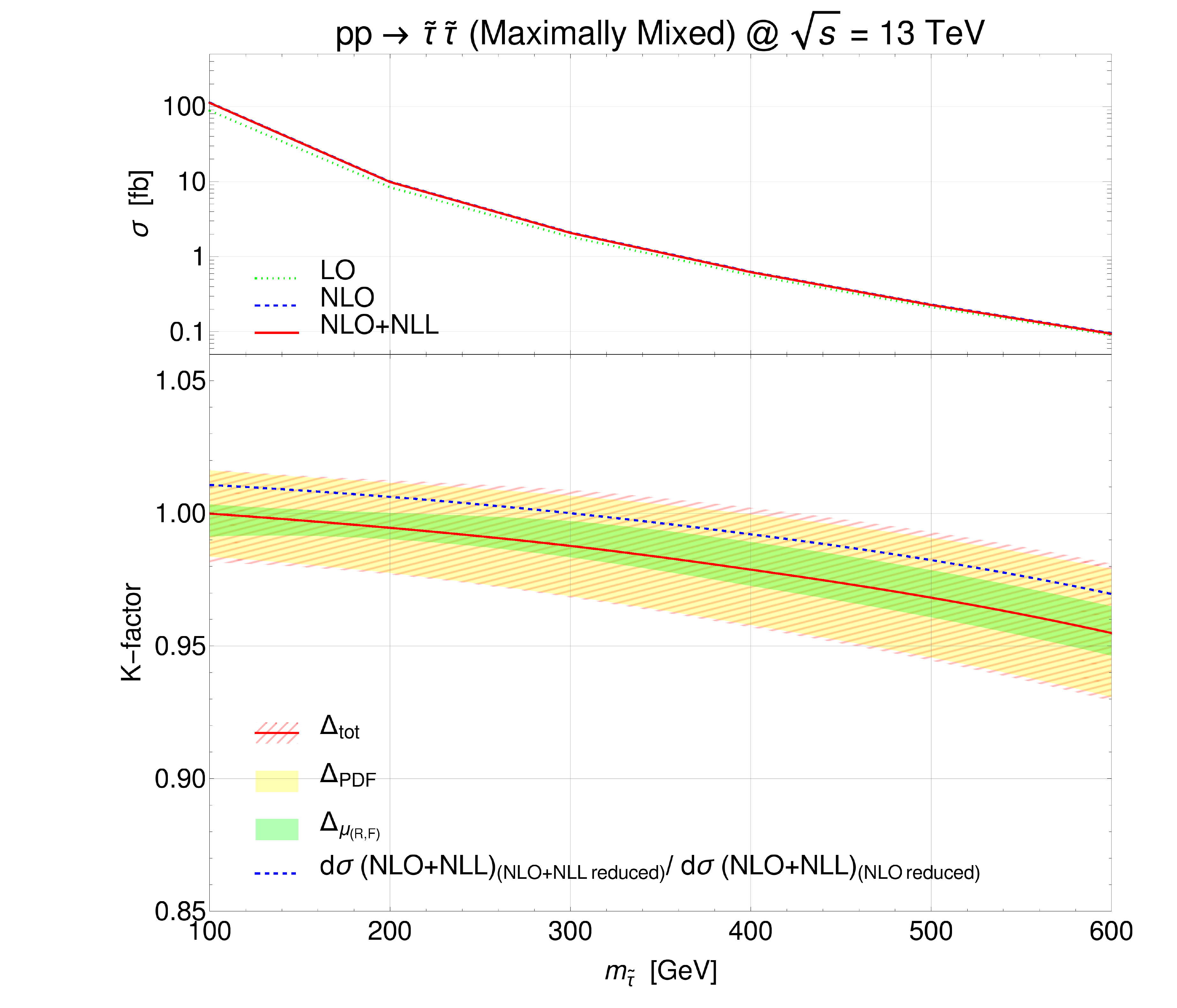}
\caption{Same as in the right plot of Fig.~\ref{fig:first_generation} for the integrated cross section of stau pair production for two mixing scenarios: totally right-handed (left) and maximally mixed (right).}
\label{fig:third_generation}
\end{center}
\end{figure}

In Fig.~\ref{fig:third_generation} we repeat the same analysis for the case of third generation sleptons.
The experimental searches for stau pair production yield less stringent limits on the mass of the SUSY particles since they also require the reconstruction of the taus in the final state~\cite{CMS-PAS-SUS-17-003}.
Moreover since in the stau sector large mixing is allowed, the experimental limits have been determined assuming different compositions of the mass eigenstates.
We show our results assuming purely right-handed staus and a maximal mixing respectively in the left and right plots of Fig.~\ref{fig:third_generation},
since the case of purely left-handed staus can be directly related to the results obtained above.
For what concerns the resulting $K$-factor we obtain similar results regardless of the stau mixing because the QCD corrections turn out to be largely independent
on the weak coupling structure of the underlying partonic cross section and the dependence on the weak couplings cancels in the ratios of Eq.~\ref{eq:K_factor}.
Similarly to the previous case, we observe a compensation between the effect of the resummation within the PDFs and the partonic matrix element.

\vspace*{-0.3cm}
\section*{Acknowledgments}
\vspace*{-0.3cm}
This work is supported by the Science and Technology Facilities Council (STFC), grant number ST/P000711/1, and by the BMBF under contract 05H15PMCCA,
and the DFG through the Research Training Network 2149 ``Strong and weak interactions from hadrons to dark matter".


\section*{References}

\begin{thebibliography}{99}

\bibitem{Ball:2017nwa} NNPDF Collaboration, \Journal{Eur. Phys. J.}{C77}{10}{2017}.

\bibitem{Accomando:2017scx} E. Accomando, J. Fiaschi, F. Hautmann, S. Moretti, {arXiv:1712.06318 [hep-ph]}

\bibitem{Dulat:2015mca} S. Dulat {\it et al}, \Journal{\PRD}{93}{3}{2016}.

\bibitem{Beenakker:1999xh} W. Beenakker {\it et al}, \Journal{\PRL}{83}{3780-3783}{1999}.

\bibitem{Fuks:2013vua} B. Fuks, M. Klasen, D.R. Lamprea, M. Rothering, \Journal{Eur. Phys. J.}{C73}{2480}{2013}.

\bibitem{Fiaschi:2018xdm} J. Fiaschi, M. Klasen, \Journal{JHEP}{1803}{094}{2018}.

\bibitem{Bonvini:2015ira} M. Bonvini {\it et al}, \Journal{JHEP}{1509}{191}{2015}.

\bibitem{Beenakker:2015rna} W. Beenakker {\it et al}, \Journal{Eur. Phys. J.}{C76}{2}{2016}.

\bibitem{CMS-PAS-SUS-17-009} CMS Collaboration, {CMS-PAS-SUS-17-009}~(2017).

\bibitem{CMS-PAS-SUS-17-003} CMS Collaboration, {CMS-PAS-SUS-17-003}~(2017).

\end{thebibliography}
\end{document}